\documentclass{jfm}
\pdfoutput=1
\usepackage{graphicx}
\usepackage[caption=false]{subfig}
\usepackage{tikz}
\usepackage{pgfplots}
\usepackage{natbib}
\usepackage{amsmath}
\pgfplotsset{compat=newest}
\definecolor{mycolor1}{rgb}{0.07843,0.07843,0.07843}% black
\definecolor{mycolor2}{rgb}{0.216 0.5 0.72}%Blue
\definecolor{mycolor3}{rgb}{0.89 0.1 0.11}% Red
\definecolor{mycolor4}{rgb}{0.3 0.69 0.29}% green
\newcommand{\blackline}{\raisebox{2pt}{\tikz{\draw[-,mycolor1,solid,line width = 0.5pt](0,0) -- (5mm,0);}}}
\newcommand{\blackdotted}{\raisebox{2pt}{\tikz{\draw[-,mycolor1,dotted,line width = 0.5pt](0,0) -- (5mm,0);}}}

\newcommand{\redline}{\raisebox{2pt}{\tikz{\draw[-,red, line width = 0.5pt](0,0) -- (5mm,0);}}}
\newcommand{\reddashed}{\raisebox{2pt}{\tikz{\draw[-,red,dashed,line width = 0.5pt](0,0) -- (5mm,0);}}}
\newcommand{\reddashdotted}{\raisebox{2pt}{\tikz{\draw[-,red,dashdotted,line width = 0.5pt](0,0) -- (5mm,0);}}}

\newcommand{\blueline}{\raisebox{2pt}{\tikz{\draw[-,mycolor2,line width = 0.5pt](0,0) -- (5mm,0);}}}
\newcommand{\purpleline}{\raisebox{2pt}{\tikz{\draw[-,mycolor4,line width = 0.5pt](0,0) -- (5mm,0);}}}
\newcommand{\mylab}[3]{\raisebox{#2}[0mm][0mm]{\makebox[0mm][l]{\hspace*{#1}#3}}}

\shorttitle{Turbulent flows over sparse canopies}
\shortauthor{A. Sharma and R. Garc{\'i}a-Mayoral}

\title{Scaling and modelling of turbulent flow over a sparse canopy}

\author{Akshath Sharma\aff{1}
\and Ricardo Garc{\'i}a-Mayoral\aff{1}
  \corresp{\email{r.gmayoral@eng.cam.ac.uk}}
 }

\affiliation{\aff{1}Dept. of Engineering, University of Cambridge,
Trumpington St., Cambridge CB2~1PZ, UK
}

\begin{document}
\maketitle
\begin{abstract}
The turbulent flow within and above a sparse canopy is investigated using direct numerical simulations. The balance of Reynolds to viscous stresses within the canopy is observed to be similar to that over a smooth wall. From this, a scaling based on their local sum is proposed. Using the conventional scaling based on the total stress, the velocity fluctuations are typically reported to be reduced within the canopy compared to smooth walls. When the proposed height-dependent scaling is used, however, the fluctuations are closer to those over smooth walls. This suggests that, in a large part, the effect of the canopy can be reduced to the modification of the local scaling, rather than to the direct interaction of the canopy elements with the turbulence. Based on this, a model is proposed that consists of a drag that acts on the mean flow alone, aiming to produce the correct scaling without modifying the fluctuations directly. This model is shown to estimate the fluctuations within the canopy better than the conventional, homogeneous-drag model. Nevertheless, homogenised methods are not able to reproduce the local effects of the canopy elements. In order to capture these, another model is proposed that applies the mean-only drag on a truncated representation of the canopy geometry. 
\end{abstract}
\section{Introduction} \label{sec:intro}
Canopies are ubiquitous in our environment, whether they be natural like forests or crops, or artificial like building clusters. The study of turbulent flows over canopies has wide-ranging applications like reducing crop loss \citep{deLangre2008}, energy harvesting \citep{Mcgarry2011}, and improving heat transfer \citep{Fazu1989}. 
On the basis of the geometry and spacing of the canopy elements, a canopy can be dense, sparse, or transitional \citep{Nepf2012,Wieringa1993}.  In the dense limit, the canopy elements are in close proximity to each other and turbulence does not penetrate within them. In the sparse limit, the spacing between canopy elements is large and the turbulent eddies can penetrate the full height of the canopy. The intermediate, transitional regime lies between these two limits. \cite{Nepf2012} proposed an approximate classification of the canopy regime based on the roughness frontal density, $\lambda_f$. % \citep{Wooding1973, Schlichting1936}. %
\cite{Nepf2012} observed that canopies are dense for $\lambda_f \gg 0.1$,  sparse for $\lambda_f \ll 0.1$, and intermediate for $\lambda_f \approx 0.1$. However, in addition to the geometric parameter, $\lambda_f$, the lengthscales of the turbulent eddies should also be considered when determining the regime of the canopy. In the present work, we study the flow within and above a sparse canopy, with particular emphasis on its scaling. The canopy has a roughness density $\lambda_f = 0.07$, with element spacings large enough to have a limited effect on the near-wall turbulent flow. We study the scaling of the flow within and above the canopy, and propose models that do not require resolving the canopy layout. Several numerical studies have modelled dense canopies using a homogeneous form drag \citep{Bailey2016, Finnigan2009}, and \cite{Busse2012} have also proposed a similar model for roughness.
\cite{Yan2017}, \cite{Bailey2013} and \cite{Yue2007} compared such models to resolving the canopy geometry. Although \cite{Bailey2013} studied spanwise-homogeneous canopies, and \cite{Yan2017} and \cite{Yue2007} heterogeneous ones, these studies showed that the homogeneous-drag approach tends to overdamp turbulent fluctuations within the canopies. \cite{Bailey2013} observed that, for large row spacing, the homogeneous model can under-predict the turbulent fluctuations by up to $50\%$. This was attributed to the lack of representation of the gaps between the canopy elements, where the fluctuations do not experience significant damping. In the present work, we propose a modification of the homogeneous drag model which only applies a drag force on the mean profile, and another that distributes the force into a reduced order representation of the canopy layout. Preliminary results from some of these simulations were presented in \cite{Sharma2018}.

The paper is organised as follows. \S\ref{sec:num_method} describes the numerical method and elaborates on the canopy geometry. The results for the resolved canopy and the scaling of turbulent fluctuations is discussed in \S\ref{sec:resolved_canopy}. \S\ref{sec:results_models} discusses the results obtained from the models. Conclusions are presented in \S\ref{sec:conclusions}. 
\begin{figure}
		\centering
           % \vspace{}
  		   	\includegraphics[width=0.7\textwidth,trim={0 1.75cm 0 1.75cm},clip]{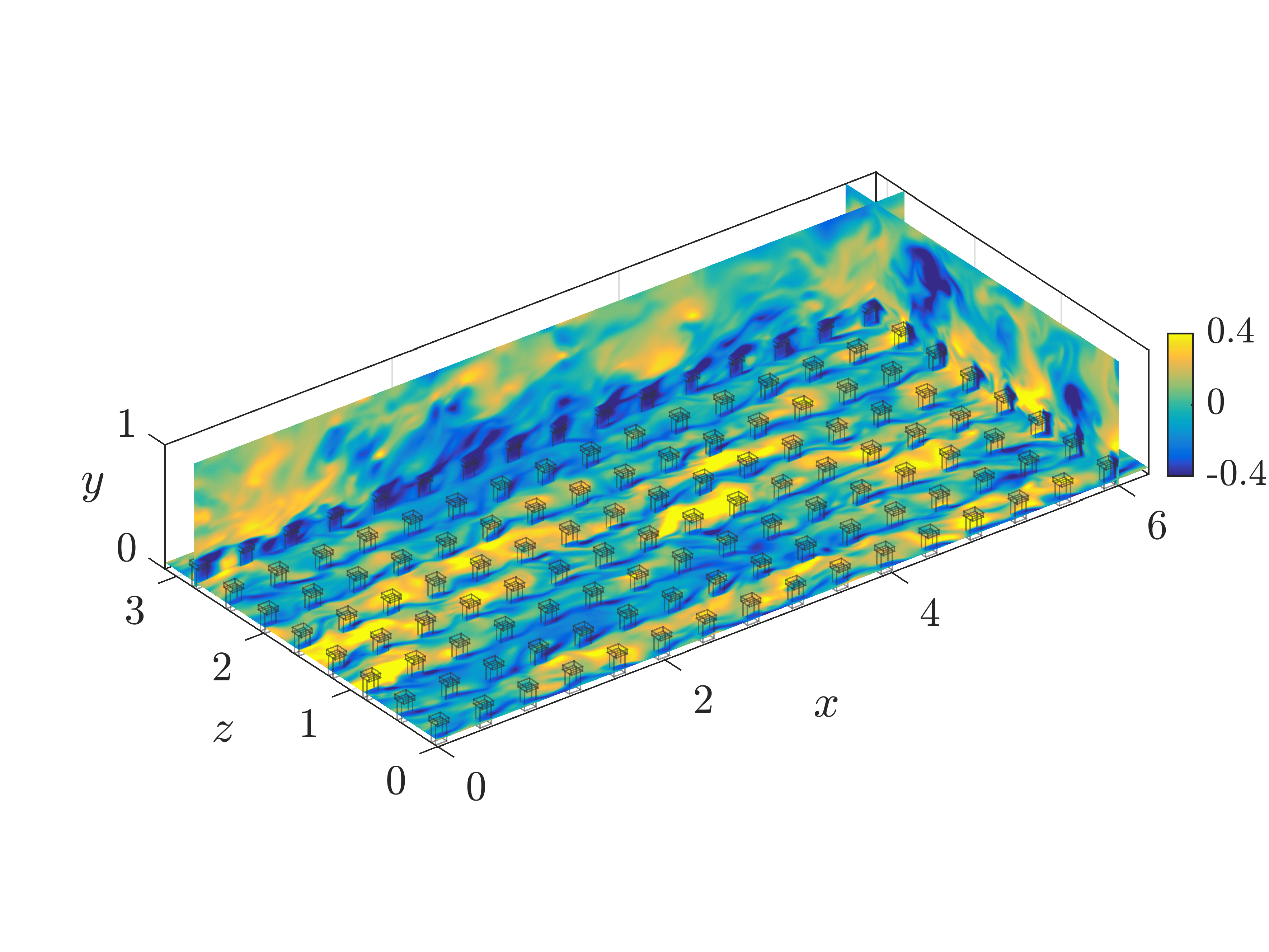}
  		    %\vspace{-1.5cm}
    		\caption{Schematic of the numerical domain with the canopy elements in wireframe. Contours represent the instantaneous fluctuating streamwise velocity.}%
	    	\label{fig:channel_schematic}	
 \end{figure}% 
 \section{Numerical simulations} \label{sec:num_method}
We conduct direct numerical simulations of an open channel with canopy elements protruding from the wall. The streamwise, wall-normal and spanwise coordinates are $x$, $y$, and $z$ respectively, and the associated velocities $u$, $v$, and $w$. The simulation domain is $2\pi \delta \times \delta \times \pi \delta$, with the channel height $\delta = 1$. The domain is periodic in the $x$ and $z$ directions. A schematic representation of the numerical domain is shown in figure \ref{fig:channel_schematic}. No-slip and impermeability conditions are applied at the bottom boundary, and free slip and impermeability at the top. The flow is incompressible, with the density set to one. The numerical method used to solve the three dimensional Navier-Stokes equations is adapted from \cite{Fairhall2018}. A Fourier spectral discretisation is used in the streamwise and spanwise directions. The wall-normal direction is discretised using second-order centred differences on a staggered grid. The wall-normal grid is stretched to give a resolution $\Delta y^+_{min} \approx 0.2$ at the wall, $y = 0$, and $\Delta y^+_{max} \approx 2$ at $y = \delta$. The time advancement is through a three-step Runge-Kutta method with a fractional step, pressure correction implementation that enforces continuity \citep{Le1991}.

The individual canopy elements are represented using an `immersed forcing' approach, similar to that used by \cite{Bailey2013} and \cite{Yue2007}. A large drag force is applied at the grid points within the canopy elements, to ensure that the velocity within is much smaller than in the surrounding points. The geometry of the canopy elements is therefore not resolved in detail, but their effect as an obstacle on the flow is represented effectively, as illustrated in figure~\ref{fig:channel_schematic}.
\begin{table}
\begin{center}
%\lineup
\begin{tabular}{lcccccccc}
%\br
\ \ \ \ \ \ \ \ \ Case &$u_\tau$ &$\Rey_\tau$ &$h^+$ &$N_c$         & $\int D^+$ &$\Delta x^+$ &$\Delta z^+$ & \\
%\mr
Smooth channel (S)    &   0.052   &538.8      & --       & --                               & --              &8.8            &4.4               & \blackdotted     \\
Resolved canopy (C)    &   0.137   &505.6      &108.6  &$16 \times 8$            &0.810         &8.27          &4.14              & \blackline\\
Homogeneous drag (H)    &   0.136   &503.3      &108.1   &--                                &0.849         &11.00        &5.50             & \blueline\\
Mean-only drag (H0)   &   0.141   &519.9      &111.7   &--                                 &0.823         &11.34        &5.67             & \redline\\
Prescribed mean flow (H0F) &   0.174  &646.0      &138.8  &--                                  &0.883         &10.57       &5.28            & \reddashed\\
Canopy-modes drag (CF)   &   0.142   &527.1      &113.3   &$16 \times 8$             &0.812         &8.62          &4.31             & \purpleline \\          
%\br
\end{tabular}%
\caption{\label{DNS_param} Simulation parameters. $u_\tau$ is the friction velocity based on the total drag in the channel, $\Rey_\tau$ is the friction Reynolds number, $h^+$ is the height of the canopy, $N_c$ and $\int D^+$ are the number of canopy elements and the drag force integrated over the canopy height respectively. $\Delta x^+$ and $\Delta z^+$ are the resolutions in the streamwise and spanwise directions.}%
\end{center}
\end{table}   
Case S is an open channel with a smooth wall. Case C has the canopy elements explicitly represented as explained above. The drag force applied is $C_{dc} {u_i} |{u_i}|$, where $C_{dc}$ is a drag coefficient and ${u_i}$ is the instantaneous local velocity in each $i$ direction. The head of the canopy elements has dimensions $l_x^+ = l_z^+ \approx 40$ in the wall-parallel directions. The base of the canopy elements has $l_x^+ \approx 40$ and $l_z^+ \approx 20$. The heights of the base and the head are $l_s^+ \approx 80$ and $l_h^+ \approx 20$, respectively. The canopy elements are in a collocated arrangement, and the spacing between the elements is $L_{x}^+ = L_{z}^+ \approx 200$. This results in a roughness density of $\lambda_f \approx 0.07$. The spanwise spacing between the canopy elements is roughly two times the width of streaks near the wall, $\lambda_{z}^+ \approx 100$, % \citep{Kline1967}, %
which implies that the canopy should be sparse from the point of view of the near-wall turbulent fluctuations as well. In case H, the presence of the canopy is modelled through a force, $C_{dh} {u_i} |{u_i}|$, applied homogeneously below the canopy tips. A second model, case H0, applies a forcing $C_{dh}  {U}  |{U}| $ in the region below the tips of the canopy, where $U(y)$ is the mean velocity profile. The drag is only applied in the streamwise direction and is only a function of $y$. A variant of the above model, case H0F, prescribes the mean velocity profile from the resolved canopy simulation, and obtains the mean-only drag required to sustain this mean profile a posteriori. Finally, the model of case CF applies a drag $C_{dh} {U} | {U} | $, like in case H0, but distributed in a reduced-order representation of the canopy elements. This consists of a 24-mode Fourier truncation of the canopy geometry.
\begin{figure}
		\centering
         \subfloat{%
  			\includegraphics{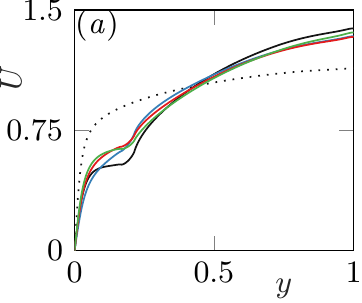}
  		}%
  		\subfloat{%
  			\includegraphics{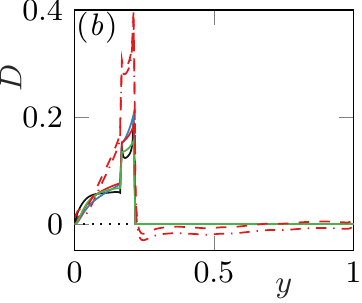}
  		}%
  	   \caption{(\textit{a}) Mean velocity profile, $U(y)$ and (\textit{b}) drag force, $D(y)$, scaled in outer units. Line styles are as defined in table~\ref{DNS_param}. \protect\reddashdotted, $D(y)$ for case H0F without accounting for the change in mean pressure gradient.}%
	    	\label{fig:U_mean}	
%	\end{figure}%
%	 \begin{figure}
		\centering
         \subfloat{%
  			\includegraphics{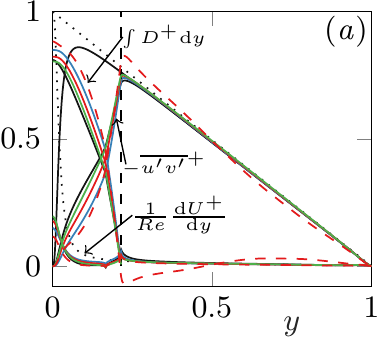}
  		}%
  		\subfloat{%
  			\includegraphics{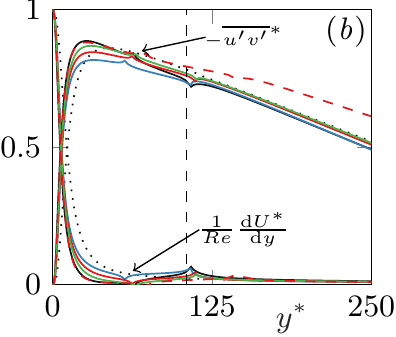}
  		}%
         \subfloat{%
  			\includegraphics{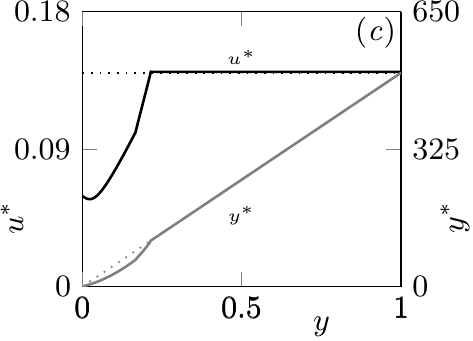}
  		}%
  	\caption{Viscous, Reynolds and drag stresses scaled with (\textit{a}) $u_{\tau}$ and (\textit{b}) $u^*$. Line styles are as in table~\ref{DNS_param}. (\textit{c})  $u^*(y)$ and $y^*(y)$ for case C.}%
	    	\label{fig:stress_profiles}	
\end{figure}%
The resolved-canopy simulation of case C was run at a constant mass flow rate, with the viscosity adjusted to obtain a friction Reynolds number based on the total stress $Re_\tau \approx 500$. The value of $C_{dc}$ was set such that further increasing it did not appreciably increase the drag experienced by the flow. The velocity within the canopy elements was then much smaller than in the surrounding points, and was essentially not reduced further by increasing the drag coefficient. An instantaneous realisation of the resulting streamwise velocity is shown in figure~\ref{fig:channel_schematic}. The subsequent simulations were run at the same mass flow rate and viscosity, with the streamwise drag coefficient, $C_{dh}$, adjusted to obtain a similar mean drag force. The wall-normal distribution of the drag force used in the simulations is shown in figure~\ref{fig:U_mean}(\textit{b}). The homogeneous drag model, case H, also required the prescription of the drag coefficients in the spanwise and wall-normal directions. They were estimated by rescaling the streamwise drag coefficient based on the relative change in the `blockage ratio' \citep{Luhar2008} of the canopy elements in the different directions, in the spirit of the method proposed by \cite{Luhar2013}. The blockage ratio in the streamwise direction is proportional to the frontal area of the canopy elements. In the wall-normal and spanwise directions this would be the top-view and the side-view areas respectively. Although this is a coarse approximation of the wall-normal drag coefficient, \cite{Busse2012} have shown that the flow is relatively insensitive to moderate changes in the wall-normal drag. The parameters for each simulation are given in table~\ref{DNS_param}.

\section{Flow through resolved canopies}\label{sec:resolved_canopy}
In this section we discuss the scaling of turbulent fluctuations for our sparse canopy, and compare them with those over a smooth wall. Over a smooth wall, the balance of stresses within the channel is obtained by averaging the momentum equations in the wall-parallel directions and time, followed by integration in $y$. This yields
\begin{eqnarray}
 \frac{\mathrm{d} P}{\mathrm{d} x} y + \tau_w & = & -\overline{u' v'} + \frac{1}{\Rey}\frac{\mathrm{d} U}{\mathrm{d} y},
  \label{eq:stress_balance_smooth}
 \end{eqnarray}
where $\tau_w$ is the wall shear stress, $\mathrm{d} P/\mathrm{d} x$ is the mean streamwise pressure gradient, $\overline{u' v'}$ is the Reynolds stress, $U$ is the mean streamwise velocity and $\Rey$ is the Reynolds number based on the bulk velocity. Particularising \eqref{eq:stress_balance_smooth} 
at $y = \delta$, we obtain the expression for the wall shear stress and the friction velocity, $u_\tau$,
\begin{eqnarray} \label{eq:utau_smooth}
 u_\tau^2  =  \tau_w  = -\delta \frac{\mathrm{d} P}{\mathrm{d} x}.
\end{eqnarray}
In our case, the stress balance also includes the drag exerted by the canopy elements,
\begin{eqnarray}\label{eq:stress_balance_canopy}
 \frac{\mathrm{d} P}{\mathrm{d} x} y + \tau_w  & = & -\overline{u' v'} + \frac{1}{\Rey}\frac{\mathrm{d} U}{\mathrm{d} y} - \int_{0}^{y} D \ \mathrm{d} y,
\end{eqnarray}
where $D$ is the canopy drag averaged in $x$, $z$ and time, and is zero for $y > h$. Equation~\eqref{eq:stress_balance_canopy} can be rewritten as
\begin{eqnarray}\label{eq:stress_balance_canopy2}
 \frac{\mathrm{d} P}{\mathrm{d} x} y + \tau_w + \int_{0}^{h} D \ \mathrm{d} y & = & -\overline{u' v'} + \frac{1}{\Rey}\frac{\mathrm{d} U}{\mathrm{d} y} + \int_{y}^{h} D \ \mathrm{d} y,
 \end{eqnarray}
so that the total drag, $\tau_w + \int_{0}^{h} D \ \mathrm{d} y$ is on the left hand side as in \eqref{eq:stress_balance_smooth}. 
From this total drag, a `global' friction velocity can be defined 
\begin{eqnarray} \label{eq:global_utau}
 u_\tau^2 = \tau_w + \int_{0}^{h} D \ \mathrm{d} y & = & -\delta \frac{\mathrm{d} P}{\mathrm{d} x}.
 \end{eqnarray} 
\noindent This is the equivalent of the smooth-wall $u_\tau$ of \eqref{eq:utau_smooth} for canopy flows.
Equation~\eqref{eq:stress_balance_canopy2}, however, shows that in canopy flows, the total stress is made up of the viscous stress, the Reynolds stress and the canopy-drag stress. Their sum, the total stress, is linear in $y$ as in smooth-wall flows. Above the canopy, the magnitude of viscous and Reynolds stresses is similar to that over smooth walls, but within the canopy the canopy drag dominates, as shown in figure~\ref{fig:stress_profiles}(\textit{a}). Let us define the sum of the viscous and Reynolds stresses as the `fluid' stress, $\tau_f$. \cite{Tuerke2013} studied smooth-wall flows with artificially forced mean profiles, and observed that $\tau_f$ provided the scale for turbulence locally. This was the case even though $\tau_f$ was not linear with $y$, as in smooth channels, due to the artificial mean-flow forcing. They defined a `local' friction velocity, $u^*$, defined by linearly extrapolating the fluid stress at each height to the wall,
\begin{equation}
{u^*(y)}^2 = \frac{\delta}{\delta - y} \tau_f(y).
\end{equation}
For a smooth, unforced channel, $u^* = u_\tau$ at every height. %A similar concept was also proposed by \cite{Hogstrom1982} for flows over urban canopies. They scaled turbulence with a local friction velocity, the square root of the magnitude of the Reynolds stress at every height, but had measurements only at heights where the contribution of the viscous stress to $\tau_f$ would be small. %
A local viscous lengthscale, $\nu/u^*$, can also be defined, and from it a height, $y^* = yu^*/\nu$. Both $u^*$ and $y^*$ are portrayed in figure~\ref{fig:stress_profiles}(\textit{c}) across the whole range of $y$ for reference. Above the canopy tips, $y > h$, $u^*$ becomes equal to the global $u_\tau$, and $y^*$ becomes $y^+$. 
\begin{figure}%
	\centering
		\subfloat{%
  		\includegraphics{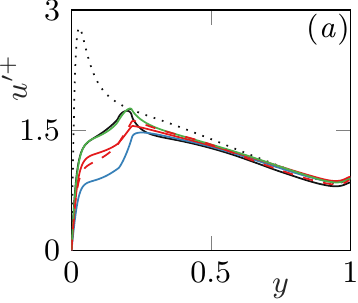}
  		}%
  		 \subfloat{%
  			\includegraphics{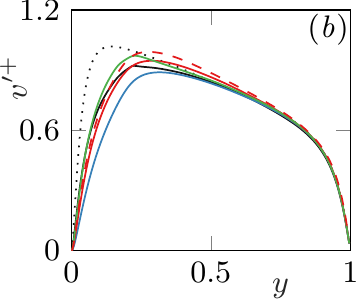}
  		}%
  		 \subfloat{%
  			\includegraphics{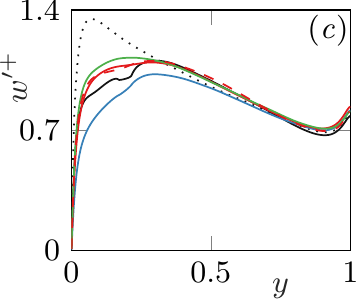}
  		}%
  		
  		\vspace*{-3.5mm}\subfloat{%
 		\hspace*{1.5mm}%
 			\includegraphics{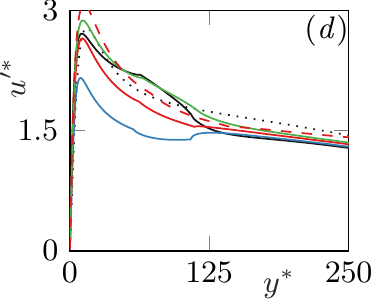}
 		}%
 		 \subfloat{%
 		\hspace*{-1mm}%
               \includegraphics{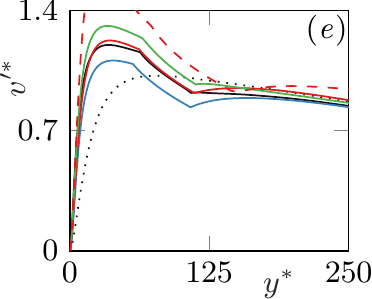}
 		}%
 		\subfloat{%
        \hspace*{-1.5mm}%
 			\includegraphics{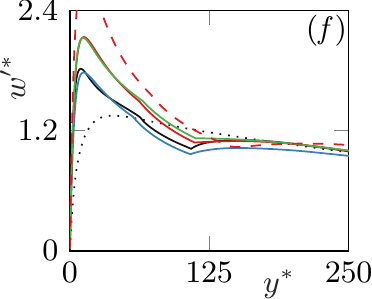}
 	    }%  		
 	    \caption{Rms velocity fluctuations scaled with (\textit{a--c}) $u_\tau$ and (\textit{d--f}) $u^*$. Line styles are as in table~\ref{DNS_param}.}%
	    	\label{fig:stats_all}%
\end{figure}%

In order to compare the local and global scalings, the terms in the stress balance within the channel scaled with $u_\tau$ and $u^*$ are portrayed in figures~\ref{fig:stress_profiles}(\textit{a}) and (\textit{b}). Figure~\ref{fig:stress_profiles}(\textit{b}) shows that even within the canopy, the ratio of viscous to Reynolds stress in $\tau_f$ remains close to that over smooth walls. The above result suggests that the effect of the canopy on the flow may be, to a large extent, a change in the local scale of turbulence, rather than a fundamental alteration of its dynamics. To further explore this, we compare the rms fluctuations scaled with both $u_\tau$ and $u^*$ in figure~\ref{fig:stats_all}. Scaling the fluctuations with $u_\tau$ produces results qualitatively similar to those reported in \cite{Bailey2013} and \cite{Yue2007}, and shows a reduction of the fluctuations within the canopy compared to a smooth wall. Scaled with $u^*$, in contrast, the streamwise fluctuations are similar to those in a smooth channel. The spanwise and wall-normal fluctuations, however, show a significant increase in magnitude within the canopy compared to a smooth wall flow, indicating a relative increase in the intensity of the crossflow. 

\begin{figure}
		\centering
     \vspace{4mm} \includegraphics[scale=1.0,trim={0mm 2.5mm 0mm 0mm},clip]{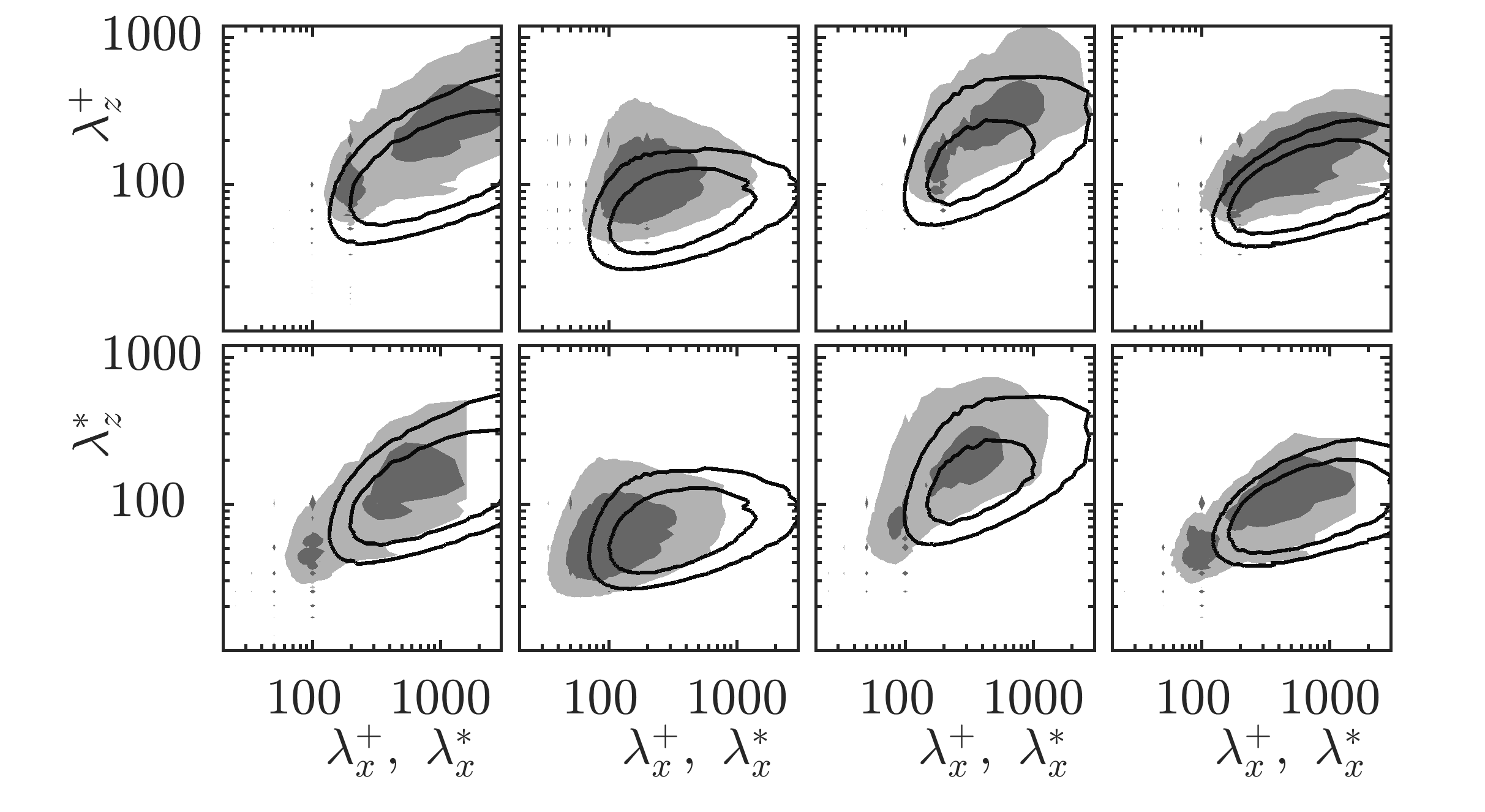}%
         \mylab{-8.65cm}{4.9cm}{(\textit{a})}%
  		\mylab{-6.6cm}{4.9cm}{(\textit{b})}%
  		\mylab{-4.55cm}{4.9cm}{(\textit{c})}%
  		\mylab{-2.5cm}{4.9cm}{(\textit{d})}%
         \mylab{-8.65cm}{2.65cm}{(\textit{e})}%
  		\mylab{-6.6cm}{2.65cm}{(\textit{f})}%
  		\mylab{-4.55cm}{2.65cm}{(\textit{g})}%
  		\mylab{-2.5cm}{2.65cm}{(\textit{h})}%
  		\mylab{-8.3cm}{5.27cm}{$k_x k_z E_{uu}$}%
   		 \mylab{-6.25cm}{5.27cm}{$k_x k_z E_{vv}$}%
   		 \mylab{-4.25cm}{5.27cm}{$k_x k_z E_{ww}$}% 	
   		 \mylab{-2.15cm}{5.27cm}{$k_x k_z E_{uv}$}% 	
		\caption{Spectral energy densities for cases C (filled contours) and S (line contours), normalised with their respective rms values at  (\textit{a--d}) $y^+ = 15$, and (\textit{e--h}) $y^* = 15$ . Wavelengths scaled in (\textit{a--d}) global units, and (\textit{e--h}) local units.\vspace*{2mm}}%The contours levels are $0.06, 0.12$ of the respective local rms value, except for (\textit{a,e}) for which they are $0.04,0.08$ times the local rms value.}%
  		\label{fig:spectra_2d_sp_cp_y15_g}
% \end{figure}
%\begin{figure}
		 \includegraphics[scale=1.0,trim={0mm 2mm 0mm 0mm},clip]{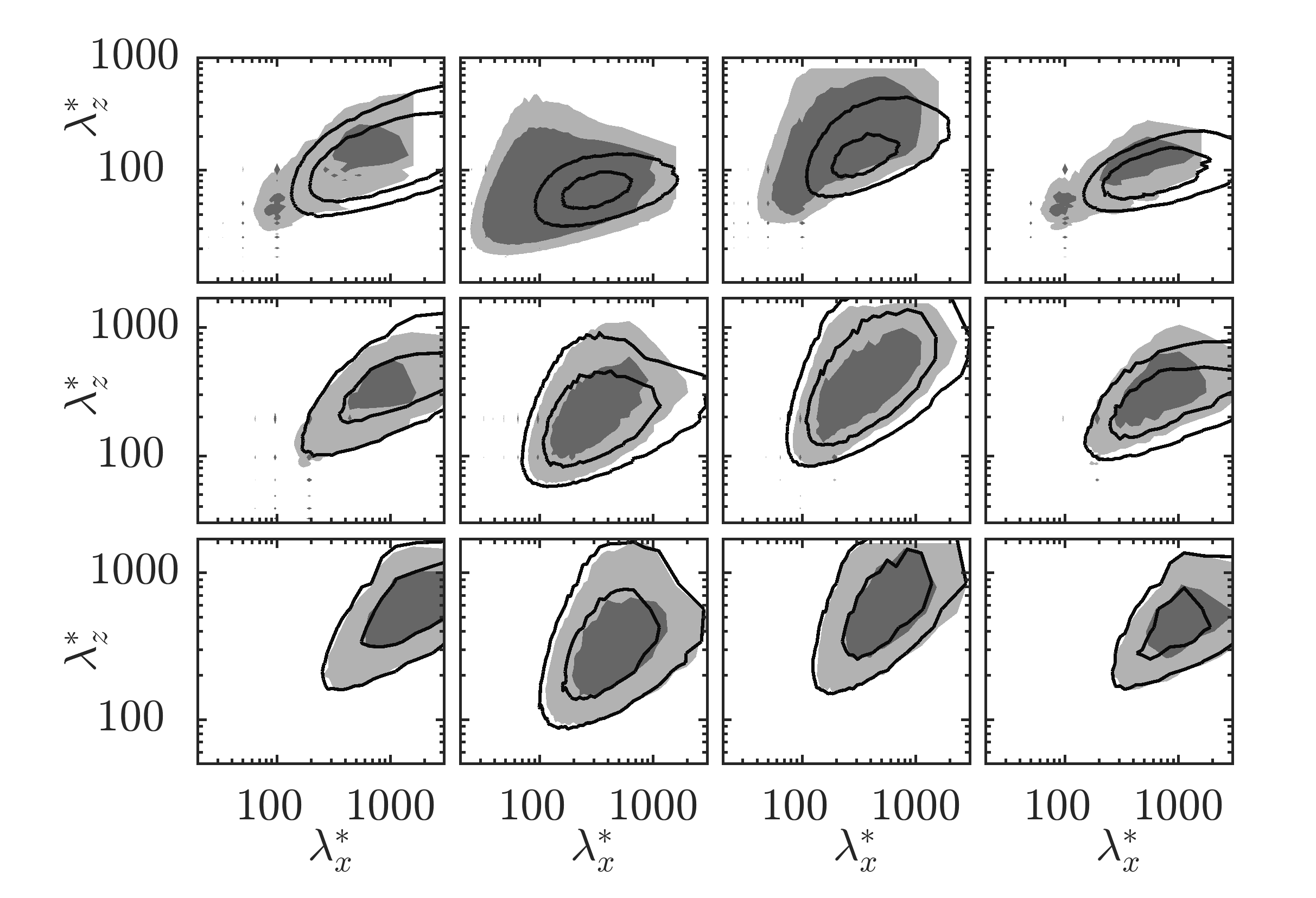}%
        \mylab{-8.7cm}{6.05cm}{(\textit{a})}%
  		\mylab{-6.65cm}{6.05cm}{(\textit{b})}%
  		\mylab{-4.6cm}{6.05cm}{(\textit{c})}%
  		\mylab{-2.55cm}{6.05cm}{(\textit{d})}%
         \mylab{-8.7cm}{4.2cm}{(\textit{e})}%
  		\mylab{-6.65cm}{4.2cm}{(\textit{f})}%
  		\mylab{-4.6cm}{4.2cm}{(\textit{g})}%
  		\mylab{-2.55cm}{4.2cm}{(\textit{h})}%
  	    \mylab{-8.7cm}{2.3cm}{(\textit{i})}%
  		\mylab{-6.65cm}{2.3cm}{(\textit{j})}%
  		\mylab{-4.6cm}{2.3cm}{(\textit{k})}%
  		\mylab{-2.55cm}{2.3cm}{(\textit{l})}%
  		\mylab{-8.3cm}{6.45cm}{$k_x k_z E_{uu}$}%
   		 \mylab{-6.25cm}{6.45cm}{$k_x k_z E_{vv}$}%
   		 \mylab{-4.25cm}{6.45cm}{$k_x k_z E_{ww}$}% 	
   		 \mylab{-2.15cm}{6.45cm}{$k_x k_z E_{uv}$}% 	
  		\caption{Spectral energy densities for cases C (filled contours) and S (line contours) normalised by $u^*$ at (\textit{a}--\textit{d}) $y^* = 15$, (\textit{e}--\textit{h}) $y^* = 105$, and (\textit{i}--\textit{l}) $y^* = 250$.\vspace*{-2mm}}% $(0.3, 0.03, 0.1, 0.06) \ u^{*2}$; $(0.15, 0.05, 0.06, 0.05)\ u^{*2}$; $(0.1, 0.04, 0.06, 0.04) \ u^{*2}$}%
   	     \label{fig:spectra_2d_sp_cp}
\end{figure}

In order to examine whether the distribution of energy across different lengthscales is similar within a canopy and over a smooth wall, we compare their spectral energy densities close to the wall in figure~\ref{fig:spectra_2d_sp_cp_y15_g}. In global units, the energy is observed to be in larger scales when compared to a smooth channel, especially in the spanwise wavelengths. In local scaling, however, the energetic wavelengths are more similar, with a greater overlap of the regions with highest intensity, particularly for $E_{uu}$ and $E_{uv}$. In addition, the canopy case exhibits a concentration of energy at the canopy harmonics --note that the canopy spacing at $y^* \approx 15$ is reduced to $L_{x}^* = L_{z}^* \approx 100$. Compared to a smooth wall, the presence of the canopy results in more energy in smaller scales and less energy in large streamwise scales when local scaling is used.
The differences in the energy distribution observed within the canopy eventually disappear above it. To illustrate this, figure~\ref{fig:spectra_2d_sp_cp} portrays the spectra within the canopy at $y^* \approx 15$, as in figure~\ref{fig:spectra_2d_sp_cp_y15_g}, just above the canopy tips at $y^* \approx 105$, and at $y^* \approx 250$. Near the tips, the canopy harmonics are weak and the smaller scales in the flow are smooth-wall-like. There is, however, a deficit of energy in large streamwise wavelengths compared to a smooth wall. This effect diminishes with height and the spectra are smooth-wall-like at $y^* \approx 250$, as shown in figures~\ref{fig:spectra_2d_sp_cp}(\textit{i--l}), delimiting the roughness sublayer. At $y^* = 15$, the scaling of the spectra with $u^*$ shows that the additional energy in $v$ and $w$ is mostly distributed along shorter streamwise wavelengths and across a wider range of spanwise wavelengths. %$E_{uu}$ shows a lack of energy in large streamwise scales, which has been associated with the shortening of streaks \citep{Flores2006}. \cite{Jimenez1999a} observed that the damping of streaks could produce more intense quasi-streamwise vortices, which is a possible cause of the increase in $v'^*$ and $w'^*$ observed.

Previous studies have noted the formation of Kelvin-Helmholtz-like rollers near the canopy tips \citep{Bailey2016, Finnigan2000}. When present, these instabilities leave a distinct footprint in $E_{vv}$ and $E_{uv}$, causing an increase in energy in a narrow range of streamwise wavelengths and for large spanwise wavelengths \citep{GG2018, Garcia-Mayoral2011}. Such a footprint is not observed in the current study. This suggests that if the instability is present over the sparse canopy, it is weak compared to the surrounding turbulent fluctuations, and is therefore masked by them.

\section{Modelling of sparse canopies} \label{sec:results_models}
The discussion in \S~\ref{sec:resolved_canopy} suggests that the canopy elements mainly affect the flow indirectly through the change in the local scale, rather than through a direct interaction. If this is the case, it would be reasonable to propose a model where the effect of the canopy is on the mean flow alone. The mean flow would, in turn, set the scale for the fluctuations. To this end, in this section we compare a conventional, homogeneous-drag model with a mean-only-drag model. In the latter case, we also discuss the results obtained from fixing the mean velocity profile and obtaining the drag force a posteriori. In addition, we also explore a model where the mean-only drag is distributed in a low-order representation of the canopy layout. 

\begin{figure}
		\centering
         \includegraphics[scale=1.0,trim={0mm 2mm 0mm 0mm},clip]{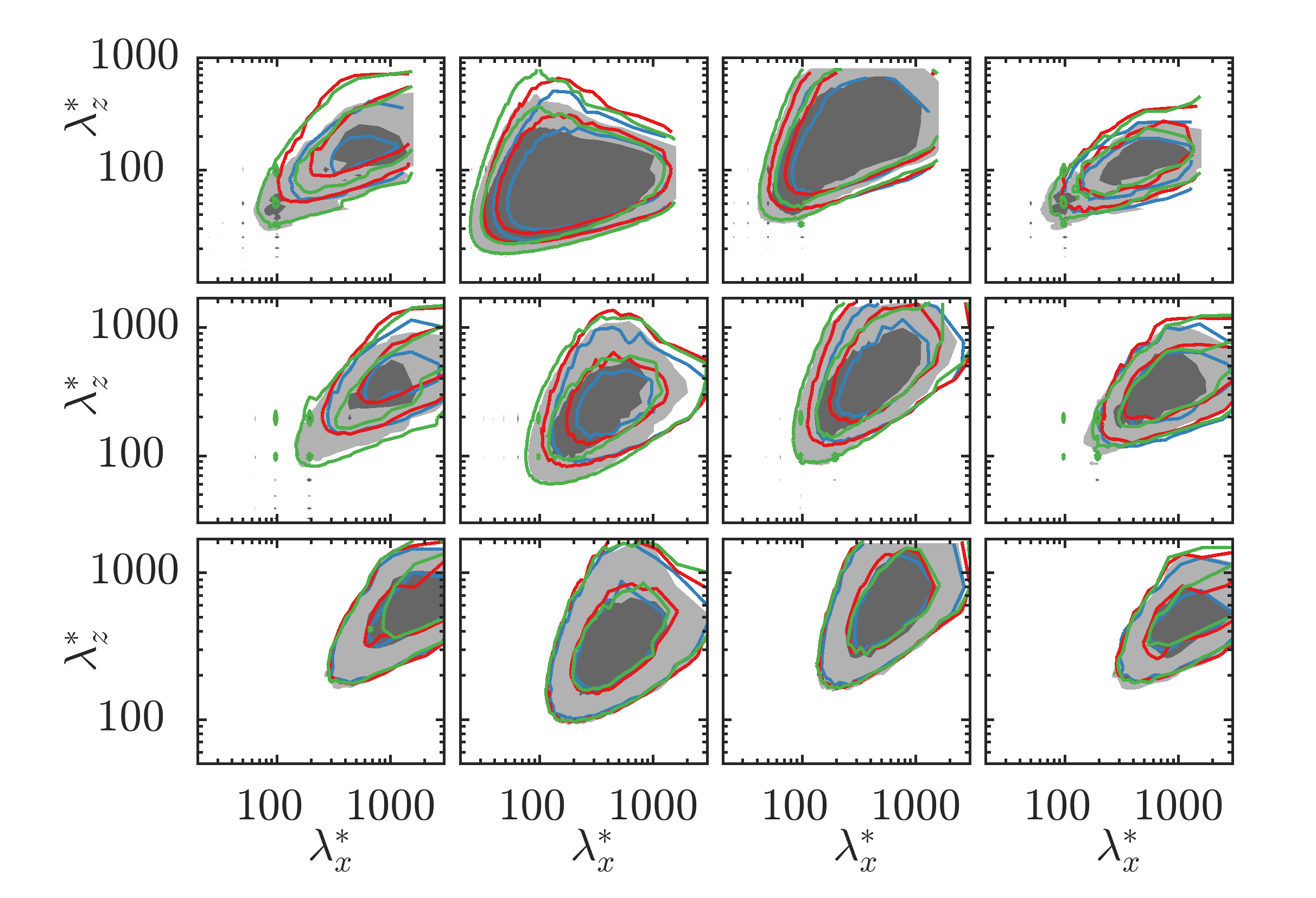}%
        \mylab{-8.7cm}{6.05cm}{(\textit{a})}%
  		\mylab{-6.65cm}{6.05cm}{(\textit{b})}%
  		\mylab{-4.6cm}{6.05cm}{(\textit{c})}%
  		\mylab{-2.55cm}{6.05cm}{(\textit{d})}%
         \mylab{-8.7cm}{4.2cm}{(\textit{e})}%
  		\mylab{-6.65cm}{4.2cm}{(\textit{f})}%
  		\mylab{-4.6cm}{4.2cm}{(\textit{g})}%
  		\mylab{-2.55cm}{4.2cm}{(\textit{h})}%
  	    \mylab{-8.7cm}{2.3cm}{(\textit{i})}%
  		\mylab{-6.65cm}{2.3cm}{(\textit{j})}%
  		\mylab{-4.6cm}{2.3cm}{(\textit{k})}%
  		\mylab{-2.55cm}{2.3cm}{(\textit{l})}%
  		\mylab{-8.3cm}{6.45cm}{$k_x k_z E_{uu}$}%
   		 \mylab{-6.25cm}{6.45cm}{$k_x k_z E_{vv}$}%
   		 \mylab{-4.25cm}{6.45cm}{$k_x k_z E_{ww}$}% 	
   		 \mylab{-2.15cm}{6.45cm}{$k_x k_z E_{uv}$}% 	
			\caption{Spectral energy densities at (\textit{a}--\textit{d}) $y^* = 15$, (\textit{e}--\textit{h}) $y^* = 105$, and (\textit{i}--\textit{l}) $y^* = 250$  normalised by $u^*$. Shaded contours represent case C and line contours are as in table~\ref{DNS_param}.\vspace*{-2mm}}%
 	\label{fig:spectra_2D_sp_md}		
 \end{figure}

When modelling the canopy through a homogeneous drag, as in case H, the rms fluctuations within the canopy are under-predicted in both global and local scaling, as shown in figure~\ref{fig:stats_all} and previously reported in the literature \citep{Yan2017, Bailey2013, Yue2007}. This is likely due to an excessive drag being directly applied on the fluctuations. Their spectral energy densities, portrayed in figure~\ref{fig:spectra_2D_sp_md}, show that the model reproduces the energy in the larger scales well. This is consistent with the fact that these scales are significantly larger than the canopy spacing, and therefore, perceive the canopy in a homogenised fashion \citep{Zampogna2016}. However, the model causes a reduction of energy in smaller scales, and obviously can not reproduce the regions of concentrated energy at the canopy wavelengths and its harmonics. This ultimately results in the under-prediction of the rms fluctuations mentioned above.

The scale-indiscriminate damping of the fluctuations by the homogeneous drag results in the overdamping of smaller scales in the flow. This should be mitigated by the mean-only drag model, case H0, which only applies a mean streamwise drag and does not directly damp the fluctuations. Compared to a homogeneous drag, this model produces rms fluctuations closer to those of the actual canopy, as shown in figure~\ref{fig:stats_all}. In global scaling, the streamwise fluctuations within the canopy are smaller than those of the resolved canopy, but in local scaling, $u'^*$ and $v'^*$ show good agreement with the resolved canopy. The peak in $w'^*$ is slightly higher, likely due to the lack of spanwise drag. The similarity in the fluctuations for the mean-only-drag model and the resolved canopy supports the idea that the sparse canopy, in large part, acts on the fluctuations through changing their scale. A comparison of the spectra, however, reveals that, despite this similarity, there are differences in the distribution of energy across different scales. Figures~\ref{fig:spectra_2D_sp_md}(\textit{a}--\textit{d}) show that, although the mean-only-drag model recovers the smaller scales to some extent, it can not capture the concentration of energy in the canopy harmonics, nor the reduction in energy in the large scales observed for the resolved canopy, as both are due to the direct interaction of the canopy elements with the flow.

The differences observed in global scaling in the fluctuations are a possible result of the mean-only drag not producing the correct mean shear. The latter is set by the mean velocity profile, which, in turn, depends on the Reynolds stress. The sum of the stresses, $\tau = -\overline{u' v'} + \frac{1}{\Rey}\frac{\mathrm{d} U}{\mathrm{d} y} + \int_{y}^{h} D \ \mathrm{d} y $, is prescribed to be linear in a channel. The latter two terms are functions of the mean velocity profile, $U(y)$. Therefore, any deviation in the Reynolds stress translates into a deviation in the mean velocity profile, and vice versa. The $U(y)$ profiles obtained from the simulations are portrayed in figure~\ref{fig:U_mean}(\textit{a}). To remove the effect of the disparity in $U(y)$ from the analysis, we conduct another simulation, case H0F, for which we fix the mean velocity profile to that of the resolved-canopy simulation, and obtain the drag force required to sustain this $U(y)$ a posteriori. The resulting drag force is shown in figure~\ref{fig:U_mean}(\textit{b}), together with the actual drag force for the resolved canopy and those from the other models. Within the canopy, the forcing required to make up for the deficit in Reynolds stress is roughly double that of the resolved canopy. Furthermore, the force is also nonzero outside the canopy. The excess forcing just above the canopy is possibly due to the Reynolds stresses generated by the canopy in the roughness sublayer. The forcing in the channel core, however, can be interpreted as an additional mean pressure gradient which, in the prescribed-$U(y)$ simulation, is combined with the drag force. We therefore obtain the actual mean pressure gradient by minimising the forcing in the channel centre. Note that, although the relative increase in drag due to the redistribution of the forcing is small, its effect on $u_\tau$ is not. The new mean pressure gradient corresponds to $\Rey_\tau \approx 650$, which is higher than the one for the simulation from which the mean velocity was prescribed, $\Rey_\tau \approx 505$. It is found, however, that the corresponding $u_\tau$ sets the scaling that recovers outer-layer similarity, as shown in the left panels of figure~\ref{fig:stats_all}. In this framework, the results are similar to those from the mean-only-drag model. In local scaling, however, we observe that the intensity of the fluctuations is significantly higher than for any of the other models, probably as a result of the balance between drag stress and $\tau_f$ being very different for this case.

The above results suggest that the mean-flow-drag models are not able to generate part of the Reynolds stress in the flow, which in case H0 is compensated for by the change in $U(y)$, and in case H0F by the drag force. It is then reasonable to conclude that the missing Reynolds stress is related to the presence of discrete canopy elements. If this is the case, some representation of the canopy geometry would be required to obtain a better approximation. Nevertheless, mean-flow models produce a more accurate representation of the flow within sparse canopies than homogeneous-drag models, which suggests that they are a more accurate representation of the effect of the canopy on the flow dynamics.
 
To introduce information on the canopy layout in the model, in case CF we distribute the drag, calculated from the mean flow as in case H0, into a reduced-order representation of the canopy elements. The representation consists of a truncation in Fourier space in $x$ and $z$ of the actual layout. In addition to capturing the most energetic scales in the flow, this model is also able to represent the concentration of energy in the canopy scales, as observed in figures~\ref{fig:spectra_2D_sp_md}(\textit{a}--\textit{d}). The drag force, although only applied in the streamwise direction, is also able to reproduce the canopy harmonics in the spectra of $v$ and $w$, reflecting the deflection of the streamwise flow around the canopy elements through continuity. The large scales in the flow are similar to those in the mean-only-drag model. The streamwise fluctuations are similar to those from the resolved canopy, both in local and global scaling, suggesting that the  difference observed in the mean-only-drag model is a result of the lack of representation of the canopy elements. The difference in the Reynolds stress, though smaller than in the mean-only-drag model, still reflects in the mean velocity profile. This likely stems from the insufficient damping on scales larger than the canopy spacing, which the model does not directly act on.
\section{Conclusions} \label{sec:conclusions}
In the present work, we have studied the turbulent flow within and above a sparse canopy. It is shown that the flow within scales with the friction velocity based on the local fluid stress, $\tau_f$, at each height, rather than the one based on the total stress. This suggests that the sparse canopy acts on the turbulent fluctuations within through a change in scale, rather than through a direct effect of the canopy elements. Based on this, a model that only applies drag on the mean flow is proposed. The model confirms that the canopy mainly acts on the flow through changing $\tau_f$. It also mitigates the excessive damping of the rms fluctuations of conventional, homogeneous-drag models, providing a better representation of the flow within the canopy. The model, however, overpredicts the energy in scales much larger than the canopy spacing, as it does not impose a drag on them like the canopy would. It also underpredicts the energy at the canopy wavelengths which is induced by the presence of the individual canopy elements. To mitigate the latter effect, an extension is proposed that distributes the drag into a reduced-order representation of the canopy elements. Although this model does not fully capture the deficit of Reynolds stress, it is able to reproduce the effect of the canopy on the smaller scales in the flow missed by the mean-only-drag model. 
\bibliographystyle{jfm}
\bibliography{jfm_ref}

\begin{thebibliography}{21}
\expandafter\ifx\csname natexlab\endcsname\relax\def\natexlab#1{#1}\fi
\def\au#1{#1} \def\ed#1{#1} \def\yr#1{#1}\def\at#1{#1}\def\jt#1{\textit{#1}}
  \def\bt#1{#1}\def\bvol#1{\textbf{#1}} \def\vol#1{#1} \def\pg#1{#1}
  \def\publ#1{#1}\def\arxiv#1{#1}\def\org#1{#1}\def\st#1{\textit{#1}}

\bibitem[Bailey \& Stoll(2013)]{Bailey2013}
{\sc \au{Bailey, B.~N.} \& \au{Stoll, R.}} \yr{2013}  \at{{Turbulence in
  Sparse, Organized Vegetative Canopies: A Large-Eddy Simulation Study}}.
  \jt{Bound.-Layer Meteorol.}  \bvol{147}~(3),  \pg{369--400}.

\bibitem[Bailey \& Stoll(2016)]{Bailey2016}
{\sc \au{Bailey, B.~N.} \& \au{Stoll, R.}} \yr{2016}  \at{{The creation and
  evolution of coherent structures in plant canopy flows and their role in
  turbulent transport}}.  \jt{J. Fluid Mech.}  \bvol{789},  \pg{425--460}.

\bibitem[Busse \& Sandham(2012)]{Busse2012}
{\sc \au{Busse, A.} \& \au{Sandham, N.~D.}} \yr{2012}  \at{{Parametric forcing
  approach to rough-wall turbulent channel flow}}.  \jt{J. Fluid Mech.}
  \bvol{712},  \pg{169--202}.

\bibitem[Fairhall \& Garc{\'i}a-Mayoral(2018)]{Fairhall2018}
{\sc \au{Fairhall, C.~T.} \& \au{Garc{\'i}a-Mayoral, R.}} \yr{2018}
  \at{Spectral analysis of the slip-length model for turbulence over textured
  superhydrophobic surfaces}.  \jt{Flow Turb. Comb.}  \bvol{100}~(4),
  \pg{961--978}.

\bibitem[Fazu \& Schwerdtfeger(1989)]{Fazu1989}
{\sc \au{Fazu, C.} \& \au{Schwerdtfeger, P, .}} \yr{1989}  \at{Flux-gradient
  relationships for momentum and heat over a rough natural surface}.  \jt{Q. J.
  Royal Meteorol. Soc.}  \bvol{115}~(486),  \pg{335--352}.

\bibitem[Finnigan(2000)]{Finnigan2000}
{\sc \au{Finnigan, J.~J.}} \yr{2000}  \at{{Turbulence in plant canopies}}.
  \jt{Annu. Rev. Fluid Mech.}  \bvol{32}~(1),  \pg{519--571}.

\bibitem[Finnigan {\em et~al.\/}(2009)Finnigan, Shaw \& Patton]{Finnigan2009}
{\sc \au{Finnigan, J.~J.}, \au{Shaw, R.~H.} \& \au{Patton, E.~G.}} \yr{2009}
  \at{Turbulence structure above a vegetation canopy}.  \jt{J. Fluid Mech.}
  \bvol{637},  \pg{387–424}.

\bibitem[Garc{\'{i}}a-Mayoral \& Jim{\'{e}}nez(2011)]{Garcia-Mayoral2011}
{\sc \au{Garc{\'{i}}a-Mayoral, R.} \& \au{Jim{\'{e}}nez, J.}} \yr{2011}
  \at{{Hydrodynamic stability and breakdown of the viscous regime over
  riblets}}.  \jt{J. Fluid Mech.}  \bvol{678},  \pg{317--347}.

\bibitem[de~Langre(2008)]{deLangre2008}
{\sc \au{de~Langre, E.}} \yr{2008}  \at{{Effects of Wind on Plants}}.
  \jt{Annu. Rev. Fluid Mech.}  \bvol{40}~(1),  \pg{141--168}.

\bibitem[Le \& Moin(1991)]{Le1991}
{\sc \au{Le, H.} \& \au{Moin, P.}} \yr{1991}  \at{An improvement of fractional
  step methods for the incompressible navier-stokes equations}.  \jt{J. Comput.
  Phys.}  \bvol{92}~(2),  \pg{369--379}.

\bibitem[Luhar \& Nepf(2013)]{Luhar2013}
{\sc \au{Luhar, M.} \& \au{Nepf, H.~M.}} \yr{2013}  \at{{From the blade scale
  to the reach scale: A characterization of aquatic vegetative drag}}.
  \jt{Adv. Water Resour.}  \bvol{51},  \pg{305--316}.

\bibitem[Luhar {\em et~al.\/}(2008)Luhar, Rominger \& Nepf]{Luhar2008}
{\sc \au{Luhar, M.}, \au{Rominger, J.} \& \au{Nepf, H.}} \yr{2008}
  \at{Interaction between flow, transport and vegetation spatial structure}.
  \jt{Environ. Fluid Mech.}  \bvol{8}~(5-6),  \pg{423}.

\bibitem[McGarry \& Knight(2011)]{Mcgarry2011}
{\sc \au{McGarry, S.} \& \au{Knight, C.}} \yr{2011}  \at{The potential for
  harvesting energy from the movement of trees}.  \jt{Sensors}  \bvol{11}~(10),
   \pg{9275--9299}.

\bibitem[Nepf(2012)]{Nepf2012}
{\sc \au{Nepf, H.~M.}} \yr{2012}  \at{{Flow and Transport in Regions with
  Aquatic Vegetation}}.  \jt{Annu. Rev. Fluid Mech.}  \bvol{44}~(1),
  \pg{123--142}.

\bibitem[G{\'o}mez-de Segura {\em et~al.\/}(2018)G{\'o}mez-de Segura, Sharma \&
  Garc{\'i}a-Mayoral]{GG2018}
{\sc \au{G{\'o}mez-de Segura, G.}, \au{Sharma, A.} \& \au{Garc{\'i}a-Mayoral,
  R.}} \yr{2018}  \at{Turbulent drag reduction using anisotropic permeable
  substrates}.  \jt{Flow Turb. Comb.}  \bvol{100}~(4),  \pg{995--1014}.

\bibitem[Sharma \& Garc{\'i}a-Mayoral(2018)]{Sharma2018}
{\sc \au{Sharma, A.} \& \au{Garc{\'i}a-Mayoral, R.}} \yr{2018}  \bt{Turbulent
  flows over sparse canopies}. ,  \vol{vol. 1001},  \pg{p. 012012}. IOP
  Publishing.

\bibitem[Tuerke \& Jim{\'{e}}nez(2013)]{Tuerke2013}
{\sc \au{Tuerke, F.} \& \au{Jim{\'{e}}nez, J.}} \yr{2013}  \at{{Simulations of
  turbulent channels with prescribed velocity profiles}}.  \jt{J. Fluid Mech.}
  \bvol{723},  \pg{587--603}.

\bibitem[Wieringa(1993)]{Wieringa1993}
{\sc \au{Wieringa, J.}} \yr{1993}  \at{Representative roughness parameters for
  homogeneous terrain}.  \jt{Bound.-Layer Meteorol.}  \bvol{63}~(4),
  \pg{323--363}.

\bibitem[Yan {\em et~al.\/}(2017)Yan, Huang, Miao, Cui \& Zhang]{Yan2017}
{\sc \au{Yan, C.}, \au{Huang, W.}, \au{Miao, S.}, \au{Cui, G.} \& \au{Zhang,
  Z.}} \yr{2017}  \at{Large-eddy simulation of flow over a vegetation-like
  canopy modelled as arrays of bluff-body elements}.  \jt{Bound.-Layer
  Meteorol.}  \bvol{165}~(2),  \pg{233--249}.

\bibitem[Yue {\em et~al.\/}(2007)Yue, Parlange, Meneveau, Zhu, van Hout \&
  Katz]{Yue2007}
{\sc \au{Yue, W.}, \au{Parlange, M.~B.}, \au{Meneveau, C.}, \au{Zhu, W.},
  \au{van Hout, R.} \& \au{Katz, J.}} \yr{2007}  \at{Large-eddy simulation of
  plant canopy flows using plant-scale representation}.  \jt{Bound.-Layer
  Meteorol.}  \bvol{124}~(2),  \pg{183--203}.

\bibitem[Zampogna \& Bottaro(2016)]{Zampogna2016}
{\sc \au{Zampogna, G.~A.} \& \au{Bottaro, A.}} \yr{2016}  \at{Fluid flow over
  and through a regular bundle of rigid fibres}.  \jt{J. Fluid Mech.}
  \bvol{792},  \pg{5--35}.

\end{thebibliography}
\end{document}